\documentclass[a4paper,12pt]{article}
\usepackage{float}
\usepackage[percent]{overpic}
\usepackage[margin=2.5cm]{geometry}
\usepackage{setspace}
\usepackage{indentfirst}
\usepackage{footmisc}
\usepackage{amsmath,amssymb,mathtools}
\usepackage{graphicx}
\usepackage{booktabs}
\usepackage{tikz}
\usetikzlibrary{arrows.meta,patterns,calc}
\usepackage{caption}
\usepackage[nosort]{cite}
\usepackage[svgnames]{xcolor}
\usepackage[colorlinks=true,allcolors=.,bookmarksnumbered=true,pdfpagemode=UseNone,pdfstartview=FitH]{hyperref}
\usepackage{microtype}
\usepackage{needspace}
\usepackage{mathrsfs}

\newcommand{\dif}{\mathrm{d}}

\numberwithin{equation}{section}
\makeatletter
\renewcommand\section{\@startsection {section}{1}{\z@}{-3.5ex \@plus -1ex \@minus -.2ex}{2.3ex \@plus.2ex}{\normalfont\large\sffamily\bfseries}}
\renewcommand\subsection{\@startsection{subsection}{2}{\z@}{-3.25ex\@plus -1ex \@minus -.2ex}{1.5ex \@plus .2ex}{\normalfont\normalsize\sffamily\bfseries}}
\renewcommand\subsubsection{\@startsection{subsubsection}{3}{\z@}{-3.25ex\@plus -1ex \@minus -.2ex}{1.0ex \@plus .2ex}{\normalfont\normalsize\sffamily\itshape}}
\makeatother

\begin{document}
\thispagestyle{empty}
\vbox{}\vspace{0.5cm}
\begin{center}
{\sffamily\LARGE Critical limits of accelerating Kerr--NUT spacetimes\\[2mm]
and the non-expanding Pleba\'nski--Demia\'nski family}\\[10mm]
{\sffamily\large Kenneth Hong}\\[4mm]
{\sffamily\slshape Department of Physics, National University of Singapore, Singapore}
\end{center}
\vspace{0.7cm}
\centerline{\sffamily\bfseries Abstract}
\bigskip
\noindent
We study degenerate-horizon limits of vacuum accelerating Kerr--NUT spacetimes with finite, nonzero NUT parameter. At cubic criticality, $a^2=m^2+\ell^2$ and $\alpha m=1$, a projective coordinate transformation relates the complete critical parent metrics by a local homothety. The cubic horizon has one finite axis and a noncompact angular end, and its near-horizon limit is locally homothetic to the swirling universe. Explicit four-dimensional coordinate transformations place the ordinary extremal, Nariai and cubic near-horizon metrics in the negative-, positive- and zero-curvature longitudinal sectors of the non-expanding Pleba\'nski--Demia\'nski family. The Nariai correspondence includes an exceptional endpoint that requires a separate Killing chart. Under a cubic-adapted near-critical scaling, fixed first-order separations between the approaching horizons do not affect the leading metric. The global Killing identifications and regular completion of the critical parent remain open.
\newpage
\section{Introduction}
The Pleba\'nski--Demia\'nski (PD) solution provides a general class of
algebraically type-D Einstein--Maxwell spacetimes with acceleration,
rotation and NUT charge \cite{PD1976}. Its original parametrization
is well suited to the algebraic classification, but the interpretation
of the parameters and the location of the horizons are less immediate.
Griffiths and Podolsk\'y reconsidered the metric and its special cases
\cite{GP2006}. More recent forms of the type-D solutions were developed
by Podolsk\'y and Vr\'atn\'y \cite{PV2023}, and the relations between
the various parametrizations were examined by Ovcharenko, Podolsk\'y
and Astorino \cite{OPA2024,OPA2025b}. In particular, their A+
parametrization exhibits the black-hole and acceleration factors of
the radial structure function explicitly. It is this form of the vacuum
accelerating Kerr--NUT metric that we shall use.

The presence of NUT charge introduces further questions concerning the
interpretation of accelerating solutions. Chng, Mann and Stelea
constructed accelerating Taub--NUT and Eguchi--Hanson solutions
\cite{CMS2006}. Podolsk\'y and Vr\'atn\'y showed that the accelerating
NUT solution of Chng, Mann and Stelea is of Petrov type I, rather than
type D \cite{PV2020}. Barrientos and Cisterna subsequently investigated
accelerating NUT solutions using Ehlers transformations and revisited
the earlier construction \cite{BC2023}. These constructions should be distinguished from the accelerating
Kerr--NUT solutions within the type-D Pleba\'nski--Demia\'nski family.

Degenerate-horizon limits of accelerating black holes have also been
investigated extensively. Dias and Lemos investigated the
extremal limits of charged, nonrotating C-metrics and obtained Nariai,
Bertotti--Robinson and anti-Nariai geometries \cite{DiasLemos2003}.
Rotation changes the limiting geometry through the coupling of the
longitudinal and azimuthal directions. The extremal near-horizon
geometry of accelerating Kerr--Newman black holes was studied by
Astorino \cite{Astorino2016}, and the accelerating Kerr--Taub--NUT
throat by Siahaan \cite{Siahaan2024}. These investigations provide
precedents for double-root limits, while the simultaneous coincidence
of three radial horizons requires a different limiting construction.

The non-expanding sector of the PD family provides a natural framework
for describing such limiting metrics. Its relation to the expanding
solutions was discussed by Griffiths and Podolsk\'y \cite{GP2006},
and a canonical treatment of the non-expanding spacetimes was given by
Podolsk\'y, Hru\v{s}ka and Griffiths \cite{PHG2018}. Their longitudinal
two-geometries admit negative, positive or zero curvature, while the
complete four-dimensional solutions can retain a nontrivial rotational
fibration. The swirling universe investigated by Astorino, Martelli
and Vigan\`o \cite{AMV2022} is a particularly relevant vacuum example.
Recently, Barrientos and Cisterna obtained this geometry from a
limiting family of rotating C-metrics in which three radial horizons
approach coincidence \cite{BC2026}. This result motivates the study
of corresponding critical configurations with nonzero NUT charge.

The local geometry of a degenerate horizon must also be distinguished
from the global spacetime obtained after imposing periodic Killing
identifications. This issue is particularly delicate for NUT solutions.
Lewandowski and Ossowski investigated nonsingular Kerr--NUT--de Sitter
spacetimes and their extensions \cite{LO2020,LO2021}, while related
work considered projectively nonsingular horizons and NUT-type
isolated-horizon structures \cite{LON2020,DRLO2023,Buk2026}.
These results provide important comparisons, but their global
constructions cannot simply be assumed to apply to the critical
accelerating family. In particular, the role of the angular endpoints
and of the Killing identification lattice must be examined in the
geometry at hand.

The rotating C-metric result of Barrientos and Cisterna thus raises
several questions for the finite-NUT type-D family: how the critical
parent geometries depend on NUT charge, how their degenerate-horizon
limits are represented within the non-expanding PD family, and what
information is retained when the horizons approach triple coincidence.
These questions concern local metric equivalence and limiting
geometries; the construction of a globally regular quotient is a
separate problem.

The paper is organized as follows. Section~\ref{sec:roots} examines
horizon degenerations of the vacuum accelerating Kerr--NUT family.
The dependence of the cubic-critical metric on NUT charge and its
near-horizon limit are studied in Sec.~\ref{sec:universal}, followed by
the horizon geometry, axial structure and global-identification questions
in Sec.~\ref{sec:geometry}. Section~\ref{sec:pd} considers the extremal,
Nariai and cubic limits in the non-expanding PD family, while
Sec.~\ref{sec:nearcritical} examines configurations approaching cubic
criticality. The discussion and conclusions are given in
Sec.~\ref{sec:conclusions}.

\section{Critical configurations of accelerating Kerr--NUT spacetimes}\label{sec:roots}
We use the vacuum A+ parametrization of the accelerating Kerr--NUT family~\cite{OPA2024}, with
$m>0$, $\alpha>0$, and vanishing electromagnetic charges and cosmological
constant. In coordinates $(t,r,x,\phi)$, the metric takes the form
\begin{align}
 \dif s^2={}&\frac{1}{\Omega^2\rho^2}
 \left[-\Delta_r(A\,\dif t-B\,\dif\phi)^2
       +\Delta_x(C\,\dif t+D\,\dif\phi)^2\right]+\frac{C_f\rho^2}{\Omega^2}
 \left(\frac{\dif r^2}{\Delta_r}+\frac{\dif x^2}{\Delta_x}\right),
 \label{eq:AKNmetric}
\end{align}
where
\begin{align}
 A(x)&=1+\alpha^2(\ell^2-a^2)x^2,&
 B(x)&=a(1+x^2)+2\ell x,\notag\\
 C(r)&=a+2\alpha\ell r+\alpha^2ar^2,&
 D(r)&=r^2+\ell^2-a^2,\label{eq:ABCD}\\
 \rho^2&=A(x)D(r)+B(x)C(r),&
 \Omega&=1-\alpha rx.\notag
\end{align}
The radial and angular functions are
\begin{align}
 \Delta_r&=(1-\alpha^2r^2)(r^2-2mr+a^2-\ell^2),
 \label{eq:dr}\\
 \Delta_x&=(1-x^2)\left[(1-\alpha mx)^2
  -\alpha^2x^2(m^2+\ell^2-a^2)\right].
 \label{eq:dx}
\end{align}
Here $m$, $a$, $\ell$ and $\alpha$ denote the mass, Kerr-like rotation,
NUT and acceleration parameters, respectively. The positive constant
$C_f$ normalizes the transverse metric and is retained explicitly in
the limiting geometries. The coordinates $t$ and $\phi$ are adapted to
the two commuting Killing fields, while $x$ is an angular coordinate.
Their global periodic identifications are not fixed at this stage.

The radial function factorizes into Kerr--NUT and acceleration
factors. For $m^2+\ell^2-a^2\geq0$, their real roots are
\begin{equation}
 r_\pm=m\pm\sqrt{m^2+\ell^2-a^2},
 \qquad r_A^\pm=\pm\alpha^{-1}.
 \label{eq:roots}
\end{equation}
We consider three degenerations involving the positive acceleration
root $r_A^+$. The first is ordinary extremality, in which the two
Kerr--NUT roots coincide below $r_A^+$:
\begin{equation}
 r_-=r_+=m<r_A^+,
 \qquad a^2=m^2+\ell^2,\qquad \alpha m<1.
 \label{eq:ext}
\end{equation}
The corresponding quadratic coefficient is
\begin{equation}
 \frac12\Delta_r''(m)=1-\alpha^2m^2>0.
 \label{eq:extcoeff}
\end{equation}
The second is the Nariai coincidence, in which the outer Kerr--NUT
root meets the positive acceleration root:
\begin{equation}
 r_-<r_+=r_A^+=r_0,\qquad
 a^2-\ell^2=2mr_0-r_0^2,\qquad r_0=\alpha^{-1}>m.
 \label{eq:nar}
\end{equation}
In this case, the quadratic coefficient is negative,
\begin{equation}
 \frac12\Delta_r''(r_0)
 =-4\left(1-\frac{m}{r_0}\right)<0.
 \label{eq:narcoeff}
\end{equation}
The two double-root branches intersect when
\begin{equation}
 r_-=r_+=r_A^+=m,\qquad
 a^2=m^2+\ell^2,\qquad \alpha m=1.
 \label{eq:cubic}
\end{equation}
The radial and angular functions then reduce to
\begin{equation}
 \Delta_r=-\frac{(r-m)^3(r+m)}{m^2},\qquad
 \Delta_x=(1-x)^3(1+x).
 \label{eq:cubicpolys}
\end{equation}
The triple root at $r=m$ has a vanishing quadratic coefficient,
whereas $r=-m$ remains a simple root. We next examine the
zero-rotation sector and parametrize the rotating cubic-critical family.
The associated limiting geometries are considered in
Sec.~\ref{sec:pd}.

\subsection{The zero-rotation obstruction}
Setting $a=0$ in the vacuum type-D A+ family gives
\begin{equation}
 r_\pm=m\pm\sqrt{m^2+\ell^2},\qquad r_+r_-=-\ell^2.
 \label{eq:tnroots}
\end{equation}
For real nonzero $\ell$, the roots satisfy $r_-<0<r_+$ and cannot
coincide. Thus cubic criticality is excluded in this zero-rotation
sector. A Nariai coincidence, however, remains possible when
$r_+=r_A^+=\alpha^{-1}$. This requires
\begin{equation}
 m=\frac{1-\alpha^2\ell^2}{2\alpha},\qquad
 \Delta_r=-\frac{1}{\alpha}(\alpha r-1)^2
 (\alpha r+1)(r+\alpha\ell^2).
 \label{eq:tnnariaifactor}
\end{equation}
For $\alpha>0$ and $\ell\ne0$, this coincidence has positive
mass precisely when $0<\alpha|\ell|<1$. The boundary
$\alpha|\ell|=1$ instead has $m=0$ and is not part of the
positive-mass Nariai branch. At the coincident positive root,
\begin{equation}
 \left.\Delta_r''\right|_{r=\alpha^{-1}}
 =-4(1+\alpha^2\ell^2)<0.
 \label{eq:tnsecond}
\end{equation}
The corresponding Nariai limiting geometry is discussed in
Appendix A of Ref.~\cite{BC2026}.
The alternative configuration in which the two black-hole roots
meet opposite acceleration horizons has $m=0$ and
$|\alpha\ell|=1$; it produces two double roots, rather than a
triple root. Barrientos and Cisterna also identify that endpoint
as locally flat \cite{BC2026}.

The obstruction applies to the zero-rotation specialization of the
type-D A+ family and should not be extended to independently
accelerating NUT solutions of more general Petrov type
\cite{CMS2006,PV2020,OPA2025b}.

\subsection{The cubic Kerr--NUT family}
For positive Kerr rotation, the cubic-critical condition
$a^2=m^2+\ell^2$ can be parametrized in terms of a rapidity
parameter $\eta$ as
\begin{equation}
 a=m\cosh\eta,\qquad \ell=m\sinh\eta,\qquad
 \alpha=m^{-1},\qquad -\infty<\eta<\infty.
 \label{eq:criticalrapidityintro}
\end{equation}
The zero-NUT rotating C-metric is recovered at $\eta=0$.
For finite $\eta$, the structure polynomials remain those of
Eq.~\eqref{eq:cubicpolys}.

We refer to the accelerating Kerr--NUT metric restricted to the
cubic-critical locus as the \emph{critical parent metric}. Its
\emph{near-horizon daughter} is obtained by a further coordinate
scaling at the triple horizon $r=m$. The critical specialization
itself involves no near-horizon limit.

Although the structure polynomials are independent of $\eta$,
the other metric coefficients are not.

\section{Critical Kerr--NUT homothety and universality}\label{sec:universal}
We examine the dependence of the complete cubic-critical metric on
the NUT rapidity before taking its near-horizon limit.

\subsection{The cubic-critical metric}
To compare the critical metrics at different values of $\eta$,
introduce dimensionless coordinates using the common scale $m$,
\begin{equation}
 R=\frac{r}{m},\qquad T=\frac{t}{m},\qquad \Phi=\phi.
 \label{eq:cubicdimensionless}
\end{equation}
Substituting the cubic-critical parametrization
(\ref{eq:criticalrapidityintro}) into the metric
(\ref{eq:AKNmetric}), we obtain
\begin{align}
 \frac{\dif s_\eta^2}{m^2}={}&
 \frac{-\Delta_R[(1-x^2)\,\dif T-b_\eta\,\dif\Phi]^2
 +\Delta_x[c_\eta\,\dif T+(R^2-1)\,\dif\Phi]^2}
 {(1-Rx)^2\Sigma_\eta}\notag\\
 &\quad{}+\frac{C_f(\eta)\Sigma_\eta}{(1-Rx)^2}
 \left(\frac{\dif R^2}{\Delta_R}+\frac{\dif x^2}{\Delta_x}\right),
 \label{eq:cubicmetric}
\end{align}
where
\begin{align}
 b_\eta(x)&=(1+x^2)\cosh\eta+2x\sinh\eta,\notag\\
 c_\eta(R)&=(1+R^2)\cosh\eta+2R\sinh\eta,\notag\\
 \Sigma_\eta(R,x)&=(1-x^2)(R^2-1)+b_\eta(x)c_\eta(R),
 \label{eq:cubicfunctions}
\end{align}
and
\begin{equation*}
 \Delta_R=-(R-1)^3(R+1),\qquad
 \Delta_x=(1-x)^3(1+x).
\end{equation*}
The two structure polynomials are independent of $\eta$, whereas
$b_\eta$, $c_\eta$ and $\Sigma_\eta$ are not. We leave the positive
transverse normalization $C_f(\eta)$ unspecified.

\subsection{Projective homothety}
The dependence of $b_\eta$ and $c_\eta$ on $\eta$ suggests
using the half-rapidity parameter
\begin{equation}
 u=\tanh\frac{\eta}{2},\qquad
 \cosh\eta=\frac{1+u^2}{1-u^2},\qquad
 \sinh\eta=\frac{2u}{1-u^2},\qquad |u|<1.
 \label{eq:rapidity}
\end{equation}
With this parametrization, consider the fractional-linear transformation
\begin{equation}
 R=\frac{Y-u}{1-uY},\qquad
 x=\frac{X-u}{1-uX},\qquad
 f(s)=\frac{1-u^2}{(1-us)^2}.
 \label{eq:cubicmobius}
\end{equation}
Here $f$ is the derivative of either fractional-linear map. Direct substitution gives
\begin{equation*}
 b_\eta(x)=f(X)b_0(X),\qquad
 c_\eta(R)=f(Y)c_0(Y).
\end{equation*}
The transformation fixes the endpoints $\pm1$ of both critical
structure polynomials.

To compare the complete four-dimensional metrics, denote their dimensionless
Killing coordinates by $(T_\eta,\Phi_\eta)$ and $(T_0,\Phi_0)$,
and choose the local identification
\begin{equation}
 \dif T_\eta=\dif T_0,\qquad
 \dif\Phi_\eta=\dif\Phi_0.
 \label{eq:cubicKillingmap}
\end{equation}
Together with (\ref{eq:cubicmobius}), this defines a local map
$F_\eta$ from the zero-NUT coordinates to those of the finite-NUT
metric. The remaining Killing coefficients satisfy
\begin{equation}
 A_\eta(x)=f(X)A_0(X),\qquad
 D_\eta(R)=f(Y)D_0(Y).
 \label{eq:cubicABCD}
\end{equation}
These identities give the transformation of the Killing one-forms and imply
\begin{equation}
 \Sigma_\eta=f(X)f(Y)\Sigma_0,\qquad
 \Omega_\eta^2=f(X)f(Y)\Omega_0^2,
 \label{eq:cubicrhoomega}
\end{equation}
with $\Omega_\eta=1-Rx$ and $\Omega_0=1-YX$. The structure
functions and coordinate differentials satisfy
\begin{align}
 \Delta_{R,\eta}(R)&=e^\eta f(Y)^2\Delta_{R,0}(Y),&
 \dif R&=f(Y)\,\dif Y,\notag\\
 \Delta_{x,\eta}(x)&=e^\eta f(X)^2\Delta_{x,0}(X),&
 \dif x&=f(X)\,\dif X.
 \label{eq:cubicpolytransform}
\end{align}

Write $g=g^{\mathrm K}+g^{\mathrm O}$ for the Killing and
radial-angular blocks of the metric. Substituting the transformation
identities into the cubic-critical metric, the projective factors
cancel within each block, giving
\begin{equation}
 F_\eta^*g_\eta^{\mathrm K}=e^\eta g_0^{\mathrm K},\qquad
 F_\eta^*g_\eta^{\mathrm O}
 =e^{-\eta}\frac{C_f(\eta)}{C_f(0)}g_0^{\mathrm O}.
\end{equation}
The two factors agree when
\begin{equation}
 C_f(\eta)=e^{2\eta}C_f(0),
 \label{eq:cubicCf}
\end{equation}
and the complete critical metric therefore satisfies
\begin{equation}
 F_\eta^*g_\eta=e^\eta g_0.
 \label{eq:cubichomothety}
\end{equation}
Hence the NUT rapidity introduces no additional local geometric
modulus of the cubic-critical parent, apart from the constant scale.

\subsection{Regular-axis normalization}
We check whether the normalization in
Eq.~\eqref{eq:cubicCf} is compatible with elementary flatness at the
finite axis $x=-1$. On the cubic locus,
$A(-1)=0$, $B(-1)=2(a-\ell)=2me^{-\eta}$, and
$\Delta_x'(-1)=8$. The degenerating Killing direction in the
original A+ coordinates is $\partial_t$. At a regular radial point,
$\rho^2(-1,r)=B(-1)C(r)$. Setting $x=-1+\delta$, we have
$\Delta_x=8\delta+O(\delta^2)$. The metric in the
$(t,\delta)$ directions has the leading form
\begin{equation}
 \dif s_{t\delta}^2
 =\frac{C_f\rho_-^2}{8\Omega_-^2\delta}\,\dif\delta^2
 +\frac{8C_-^2\delta}{\Omega_-^2\rho_-^2}\,\dif t^2
 +O(1)\,\dif\delta^2+O(\delta^2)\,\dif t^2,
 \label{eq:axislocalexpansion}
\end{equation}
where $\rho_-^2=B(-1)C_-$ and $C_-=C(r)$. To avoid a conical
singularity at $x=-1$, elementary flatness requires $t$ to have period
\begin{equation}
 \Delta t_{\mathrm{axis}}
 =\frac{4\pi\sqrt{C_f}\,|B(-1)|}{|\Delta_x'(-1)|}
 =\pi m e^{-\eta}\sqrt{C_f(\eta)}.
 \label{eq:axisperiod}
\end{equation}
With Eq.~\eqref{eq:cubicCf}, this becomes
\begin{equation}
 \Delta t_{\mathrm{axis}}=\pi m\sqrt{C_f(0)},
 \label{eq:axisperiodfixed}
\end{equation}
independent of $\eta$. The homothety normalization therefore makes the finite-axis period
independent of $\eta$. Identifying the orbits of $\partial_t$ with
this period ensures elementary flatness there.

\subsection{The universal cubic daughter}
Consider the near-horizon limit of the critical parent at its
triple horizon $r=m$. Barrientos and Cisterna showed that this limit
of the zero-NUT rotating C-metric gives the swirling universe
\cite{BC2026}. We use the parent homothety to determine the
corresponding daughter for arbitrary finite NUT rapidity. Choose
$C_f(0)=1/2$ for the reference representative, corresponding to the
regular-axis normalization of the rotating C-metric. In these conventions, the reference limiting metric is
\cite{AMV2022,BC2026}
\begin{align}
 \dif s_{\mathrm{swirl}}^2={}&
 (1+j^2\varrho^4)
 (-\dif T^2+\dif\varrho^2+\dif Z^2)+\frac{\varrho^2}{1+j^2\varrho^4}
 (\dif\varphi+4jZ\,\dif T)^2,
 \qquad j=\frac{2}{m^2},
 \label{eq:swirldaughter}
\end{align}
the swirling universe of Astorino, Martelli and Vigan\`o
\cite{AMV2022}.

To exhibit the reference limit explicitly, express the zero-NUT A+ metric in the rotating C-metric
Killing coordinates $(t_{\mathrm C},\phi_{\mathrm C})$.
At $a=m$, $\ell=0$ and $\alpha=m^{-1}$, the required local change of
Killing basis is
\begin{equation}
 \dif t_{\mathrm A}=\frac{\dif t_{\mathrm C}}{\sqrt2}
 -\sqrt2m\,\dif\phi_{\mathrm C},\qquad
 \dif\phi_{\mathrm A}=-\frac{\dif t_{\mathrm C}}{\sqrt2m}.
 \label{eq:cubicKillingreference}
\end{equation}
Together with $C_f(0)=1/2$, this brings the zero-NUT A+ metric to
the rotating C-metric convention used in the reference scaling.
Writing $x=\cos\theta$ and using the radial function $\Delta_r$ of Sec.~2, introduce ingoing coordinates by
\begin{equation}
 \dif v=\dif t_{\mathrm C}+\frac{r^2+m^2}{\Delta_r}\,\dif r,
 \qquad
 \dif\psi=\dif\phi_{\mathrm C}+\frac{m}{\Delta_r}\,\dif r.
 \label{eq:cubicIngoingReference}
\end{equation}
The cubic near-horizon scaling in this regular chart is
\begin{equation}
 r=m(1+\varepsilon\widehat R),\qquad
 v=\frac{2m}{\varepsilon}V,\qquad
 \psi=\widehat\Phi+\frac{V}{\varepsilon},
 \label{eq:cubicReferenceScaling}
\end{equation}
with $(V,\widehat R,\theta,\widehat\Phi)$ fixed as
$\varepsilon\to0$. Here $\widehat R$ is a near-horizon coordinate,
not the dimensionless parent coordinate $R$ of Sec.~3.1.
The limiting metric is brought to the swirling form by
\begin{equation}
 \varrho=\frac{m}{\sqrt2}\cot\frac{\theta}{2},\qquad
 V=\frac{Z+T}{m},\qquad
 \widehat R=\frac{Z-T}{m},\qquad
 \widehat\Phi=\frac{\varphi}{4}
 -\frac{Z^2-T^2-2ZT}{2m^2}.
 \label{eq:cubicSwirlCoordinates}
\end{equation}
Let $H_\varepsilon$ denote the map from the near-horizon
coordinates to the zero-NUT A+ coordinates obtained by combining
\eqref{eq:cubicKillingreference}--\eqref{eq:cubicReferenceScaling},
and let $W$ be the transformation
\eqref{eq:cubicSwirlCoordinates} from swirling to near-horizon
coordinates. The complete reference map is therefore
$S_\varepsilon=H_\varepsilon\circ W$, and the known zero-NUT
limit gives
$\lim_{\varepsilon\to0}S_\varepsilon^*g_0=g_{\mathrm{swirl}}$.
For finite $\eta$, the projective map remains nonsingular at
$Y=1$, since
\begin{equation}
 \left.\frac{\partial R}{\partial Y}\right|_{Y=1}
 =\frac{1+u}{1-u}>0.
 \label{eq:cubicjacobian}
\end{equation}
Applying the reference scaling to the homothety
\eqref{eq:cubichomothety} gives
\begin{equation}
 \lim_{\varepsilon\to0}
 S_\varepsilon^*F_\eta^*g_\eta
 =e^\eta g_{\mathrm{swirl}}.
 \label{eq:cubictransport}
\end{equation}
Thus, for arbitrary finite NUT rapidity, the cubic near-horizon limit
of the accelerating Kerr--NUT metric is locally homothetic to the
swirling universe.

\section{Geometry and global structure of the cubic-critical Kerr--NUT spacetime}\label{sec:geometry}
We examine the horizons, axial structure and boundary loci of the critical parent.
The homothety of Sec.~3.2 allows these questions to be studied using the zero-NUT representative.

\subsection{Critical horizons and their extensions}
At cubic criticality, $r=m$ is a triple root of $\Delta_r$, while
$r=-m$ remains simple. In the A+ coordinates,
\begin{equation}
 g^{rr}=\frac{\Omega^2\Delta_r}{C_f\rho^2}.
 \label{eq:generalcriticalnormal}
\end{equation}
On regular patches with $C_f>0$, $\rho^2>0$ and $\Omega\ne0$,
the radial normal changes causal character across the cubic horizon,
despite its vanishing surface gravity.

To establish local extendibility, use the zero-NUT rotating C-metric
coordinates introduced in Sec.~3.4. The ingoing transformation
\eqref{eq:cubicIngoingReference} gives the complete critical metric
\begin{align}
 \dif s_0^2=\frac{1}{\Omega_0^2}\bigg[&
 -\frac{\Delta_r}{\Sigma_0}
 (\dif v-m\sin^2\theta\,\dif\psi)^2
 +2(\dif v-m\sin^2\theta\,\dif\psi)\,\dif r\notag\\
 &\quad{}+\frac{\Sigma_0}{P_0}\,\dif\theta^2
 +\frac{P_0\sin^2\theta}{\Sigma_0}
 \big(m\,\dif v-(r^2+m^2)\,\dif\psi\big)^2\bigg],
 \label{eq:criticalingoingmetric}
\end{align}
where $\Sigma_0=r^2+m^2\cos^2\theta$,
$P_0=(1-\cos\theta)^2$ and
$\Omega_0=1-(r/m)\cos\theta$.
The absence of inverse powers of $\Delta_r$ shows that the metric
extends smoothly across the regular portions of both radial horizons;
the corresponding outgoing chart covers the complementary branches.
The negative acceleration horizon at $r=-m$ is nondegenerate,
whereas the cubic horizon at $r=m$ has vanishing surface gravity
and admits no regular bifurcation surface.

The homothety transfers these local extensions to the finite-NUT
metrics. At the two radial horizons, the projective map obeys
\begin{equation}
 \left.\frac{\partial R}{\partial Y}\right|_{Y=1}
 =\frac{1+u}{1-u},\qquad
 \left.\frac{\partial R}{\partial Y}\right|_{Y=-1}
 =\frac{1-u}{1+u},\qquad |u|<1.
 \label{eq:criticalhorizonjacobians}
\end{equation}
For finite $\eta$, both derivatives are finite and nonzero. The local extensions across the regular portions of $r=\pm m$ therefore carry over to the finite-NUT metrics.

\Needspace{12\baselineskip}
\subsection{Horizon geometry and axial structure}
Consider a spatial section of the cubic horizon. At $R=1$, the horizon generator is $\partial_\Phi$.
The metric induced on a section of constant $\Phi$ follows from
Eq.~\eqref{eq:cubicmetric}:
\begin{align}
 \dif s_H^2={}&\frac{2m^2e^\eta\Delta_x}
 {(1-x)^2b_\eta(x)}\,\dif T^2
 +\frac{2m^2e^\eta C_f(\eta)b_\eta(x)}
 {(1-x)^2\Delta_x}\,\dif x^2,
 \label{eq:criticalhorizonmetric}
\end{align}
where $\Delta_x=(1-x)^3(1+x)$.
At $x=-1$, the simple zero of $\Delta_x$ gives a finite-distance
axis, with the elementary-flatness period already determined in
Sec.~3.3. At the opposite endpoint, $x\to1^-$, the radial
coefficient behaves as
\begin{equation}
 g_{xx}\sim\frac{2m^2e^{2\eta}C_f(\eta)}{(1-x)^5}.
 \label{eq:criticalhorizondistance}
\end{equation}
The proper distance to this endpoint diverges, since
$\int^1(1-x)^{-5/2}\,\dif x$ is infinite. Thus the
cubic-horizon section has one finite axis and one noncompact end.

Let $\gamma_\eta$ denote the induced metric on a spatial section of
the finite-NUT cubic horizon, given in Eq.~\eqref{eq:criticalhorizonmetric},
and $K_\eta$ its Gaussian curvature. We denote the corresponding
zero-NUT quantities by $\gamma_0$ and $K_0$. Restricting the projective
homothety of Sec.~3.2 to corresponding horizon sections gives
\begin{equation}
 F_\eta^*\gamma_\eta=e^\eta\gamma_0,\qquad
 F_\eta^*K_\eta=e^{-\eta}K_0.
 \label{eq:invariant-horizon-scaling}
\end{equation}
Thus the intrinsic horizon metrics are locally homothetic.

The limiting kernel directions of the unrescaled Killing block
at $x=\pm1$ are
\begin{equation}
 v_\pm=B(\pm1)\partial_t+A(\pm1)\partial_\phi,
 \qquad A(\pm1)=0,\quad B(\pm1)=2me^{\pm\eta}.
 \label{eq:criticalrodvectors}
\end{equation}
Both limiting kernel vectors are proportional to $\partial_t$,
although their normalizations differ. The finite axis at $x=-1$ has the period found in Sec.~3.3.
Since $x=1$ is at infinite intrinsic distance, its formal kernel direction does not define a second regular axis or impose another elementary-flatness condition.

\subsection{Curvature singularity, chronology and conformal boundary}
The projective map relates the singularity and boundary loci on regular patches.
The image of the reference ring singularity lies at $\Sigma_\eta=0$, with
\begin{equation}
 \frac{r_{\rm sing}}m=x_{\rm sing}
 =-\tanh\frac\eta2.
 \label{eq:singularitymap}
\end{equation}
Indeed, at $R=x=-u$ the critical functions satisfy
$b_\eta(-u)=c_\eta(-u)=1-u^2$, so that
$\Sigma_\eta(-u,-u)=0$. Under the projective transformation, this
locus corresponds to $Y=X=0$ in the zero-NUT representative. Its coordinate location therefore varies with $\eta$.

For a specified periodic Killing field $\xi_0$ in the reference
geometry, let $\xi_\eta=(F_\eta)_*\xi_0$ denote the corresponding
finite-NUT field. The boundary of the region containing timelike
orbits of this field is determined by
$g_\eta(\xi_\eta,\xi_\eta)=0$. Since the homothety has a positive
constant scale factor, the sign of the norm is preserved. For the transported periodic identification, the corresponding chronology boundaries are therefore projectively related. This does not classify closed timelike curves arising from other identifications.

The conformal factor $\Omega=1-Rx$ transforms according to
\begin{equation}
 1-Rx=\frac{(1-u^2)(1-YX)}{(1-uY)(1-uX)},
 \qquad u=\tanh\frac\eta2.
 \label{eq:conformalmap}
\end{equation}
Thus the conformal-boundary loci $\Omega=0$ correspond wherever
the projective map is regular.

\subsection{Global identifications and limitations}
The elementary-flatness condition at the single finite axis fixes its period (Sec.~3.3), but does not determine additional Misner-type identifications.

Global regularity and Killing-bundle constructions for NUT
spacetimes have been investigated in nonextremal and extremal
settings \cite{LO2020,LON2020,LO2021,DRLO2023,Buk2026}.
Those results do not settle the global completion of the present noncompact cubic-critical geometry, whose horizon structure and Killing identifications require separate treatment.

The existence of a globally regular Killing quotient or maximal
analytic extension of the cubic-critical Kerr--NUT spacetime thus
remains open.

\section{Degenerate limits and the non-expanding PD family}\label{sec:pd}
We now compare the near-horizon limits of Sec.~\ref{sec:roots}
with the non-expanding Pleba\'nski--Demia\'nski (PD) family.
The comparison uses the complete four-dimensional metrics, including
the transverse and rotational terms.

\subsection{The non-expanding PD solution}
For these identifications we use the canonical vacuum non-expanding
PD metric of Podolsk\'y, Hru\v{s}ka and Griffiths \cite{PHG2018},
\begin{align}
 \dif s^2={}&(p^2+\gamma^2)
 \left[-Q(q)\,\dif t^2+\frac{\dif q^2}{Q(q)}\right]
 +\frac{p^2+\gamma^2}{P(p)}\,\dif p^2+\frac{P(p)}{p^2+\gamma^2}
 (\dif y+2\gamma q\,\dif t)^2,
 \label{eq:npd}
\end{align}
where, in vacuum,
\begin{equation}
 Q(q)=\epsilon_0-\epsilon_2q^2,\qquad
 P(p)=-\epsilon_2\gamma^2+2np+\epsilon_2p^2.
 \label{eq:pdpolys}
\end{equation}
The repeated principal null congruences of this vacuum type-D metric are
non-expanding, unlike those of the accelerating Kerr--NUT parent. The
transition between these optical classes is effected by a singular
coordinate and parameter limit, not by setting the acceleration to zero
\cite{GP2006,PHG2018}. The longitudinal two-metric in square brackets
has scalar curvature $R^{(2)}=2\epsilon_2$. Its negative, positive and
zero curvature classes are represented, respectively, by
$\epsilon_2=-1,+1,0$. These labels describe the longitudinal
geometry, not a direct-product decomposition of the full metric: the
last term in Eq.~\eqref{eq:npd} generally defines a non-trivial
Killing fibration. The parameter $n$ is a canonical PD parameter and
should not be identified with the physical NUT parameter $\ell$.

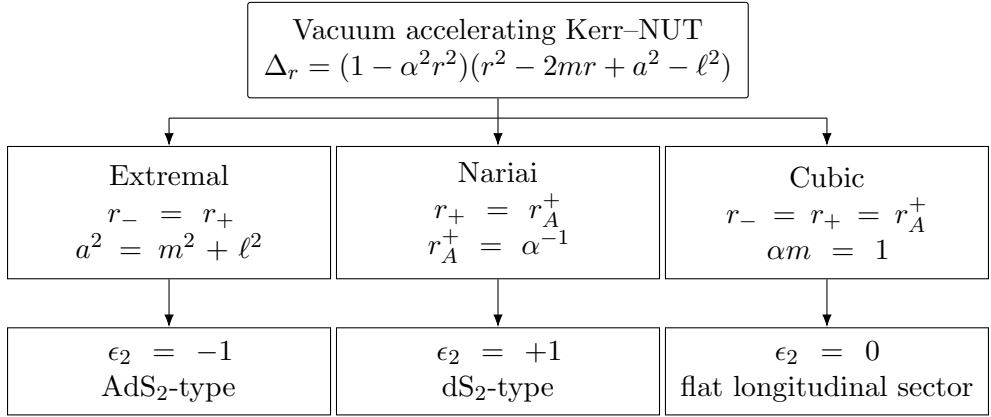
\begin{figure}[H]
\centering
\begin{tikzpicture}[>=Latex, font=\small,
 parent/.style={draw,rounded corners=1pt,align=center,inner sep=6pt},
 branch/.style={draw,align=center,inner sep=5pt,text width=0.245\textwidth,
 minimum height=1.75cm},
 daughter/.style={draw,align=center,inner sep=5pt,text width=0.245\textwidth,
 minimum height=1.15cm}]
\node[parent] (seed) at (0,3.65) {Vacuum accelerating Kerr--NUT\\
 $\Delta_r=(1-\alpha^2r^2)(r^2-2mr+a^2-\ell^2)$};
\node[branch] (ext) at (-4.35,1.5) {Extremal\\$r_-=r_+$\\$a^2=m^2+\ell^2$};
\node[branch] (nar) at (0,1.5) {Nariai\\$r_+=r_A^+$\\$r_A^+=\alpha^{-1}$};
\node[branch] (cub) at (4.35,1.5) {Cubic\\$r_-=r_+=r_A^+$\\$\alpha m=1$};
\node[daughter] (e) at (-4.35,-0.6) {$\epsilon_2=-1$\\AdS$_2$-type};
\node[daughter] (n) at (0,-0.6) {$\epsilon_2=+1$\\dS$_2$-type};
\node[daughter] (c) at (4.35,-0.6) {$\epsilon_2=0$\\flat longitudinal sector};
\coordinate (fork) at (0,2.75);
\draw (seed.south) -- (fork);
\draw (ext.north |- fork) -- (cub.north |- fork);
\draw[->] (ext.north |- fork) -- (ext.north);
\draw[->] (fork) -- (nar.north);
\draw[->] (cub.north |- fork) -- (cub.north);
\draw[->] (ext.south) -- (e.north);
\draw[->] (nar.south) -- (n.north);
\draw[->] (cub.south) -- (c.north);
\end{tikzpicture}
\caption{The three degenerate-horizon limits of the vacuum accelerating
Kerr--NUT family and the corresponding longitudinal-curvature sectors
of the non-expanding PD family. The cubic branch additionally obeys
$a^2=m^2+\ell^2$. The exceptional Nariai endpoint belongs to the
same $\epsilon_2=+1$ sector.}
\label{fig:pdbranches}
\end{figure}

Figure~\ref{fig:pdbranches} shows the three horizon coincidences
and their corresponding PD sectors. We use a common double-root
limit for the extremal and Nariai branches, and the swirling metric
of Sec.~\ref{sec:universal} for the cubic branch.

\subsection{Double-root limiting metric}
The extremal and Nariai branches both involve a double root of
$\Delta_r$, but with quadratic coefficients of opposite signs.
To treat both cases together, let $r_h$ denote a double root of
$\Delta_r$, and define
\begin{equation}
 C_h=C(r_h),\quad D_h=D(r_h),\quad
 \lambda_h=\tfrac12\Delta_r''(r_h),\quad
 W_h=C'(r_h)D_h-D'(r_h)C_h.
 \label{eq:doubledata}
\end{equation}
For $C_h\ne0$ and $\lambda_h\ne0$, introduce the co-rotating
coordinates
\begin{equation}
 r=r_h+\varepsilon R,\qquad
 t=D_h\frac{\tau}{\varepsilon}+\Phi,\qquad
 \phi=-C_h\frac{\tau}{\varepsilon}.
 \label{eq:doublescale}
\end{equation}
Using $\rho_h^2=A(x)D_h+B(x)C_h$ and
$\Omega_h=1-\alpha r_hx$, the finite local limit is
\begin{align}
 \dif s_h^2={}&\frac{\rho_h^2}{\Omega_h^2}
 \left[-\lambda_hR^2\,\dif\tau^2
 +\frac{C_f}{\lambda_h}\,\frac{\dif R^2}{R^2}
 +\frac{C_f}{\Delta_x}\,\dif x^2\right]+\frac{\Delta_x}{\Omega_h^2\rho_h^2}
 (C_h\,\dif\Phi+W_hR\,\dif\tau)^2.
 \label{eq:generaldouble}
\end{align}
For both branches considered below, the exact horizon functions give
$W_h=-2mC_h$. Equation~\eqref{eq:generaldouble} follows from the
complete A+ metric, including the transverse factor $C_f$; it is a
local double-root limit and does not impose a global angular quotient.

\subsection{Ordinary extremal branch}
The extremal throat has been studied in Ref.~\cite{Siahaan2024}.
To express it in canonical non-expanding PD coordinates, define
\begin{equation}
 q_0=\alpha m,\quad \lambda=1-q_0^2,\quad
 C_m=a(1+q_0^2)+2\ell q_0,\quad
 H=a+\ell q_0,\quad J=\ell+aq_0.
 \label{eq:extconstants}
\end{equation}
Here $D(m)=0$ and $\rho_m^2=C_mB(x)$. The transverse coordinate
\begin{equation}
 p_0=\frac{Hx+J}{1-q_0x}
 \label{eq:extp0}
\end{equation}
satisfies the exact identities
\begin{align}
 \frac{\rho_m^2}{(1-q_0x)^2}&=p_0^2+m^2,
 &\frac{\dif p_0}{\dif x}&=\frac{C_m}{(1-q_0x)^2},\notag\\
 \frac{C_m^2\Delta_x}{(1-q_0x)^4}
 &=\lambda(m^2-p_0^2)+2Np_0,
 &N&=\ell(1+q_0^2)+2aq_0.
 \label{eq:extidentities}
\end{align}
To express the full metric in canonical PD form, we introduce the following coordinates and parameters:
\begin{align}
 p&=\sqrt{\frac{C_f}{\lambda}}\,p_0,
 &q&=R,
 &t&=\frac{\lambda}{\sqrt{C_f}}\tau,
 &y&=-\sqrt{\lambda}\,\Phi,\notag\\
 \gamma&=m\sqrt{\frac{C_f}{\lambda}},
 &n&=\frac{N\sqrt{C_f}}{\lambda^{3/2}},
 &\epsilon_0&=0,
 &\epsilon_2&=-1.
 \label{eq:extdictionary}
\end{align}
Under this transformation, the metric \eqref{eq:generaldouble} takes the canonical form \eqref{eq:npd}, with
\begin{equation}
 Q(q)=q^2,\qquad P(p)=\gamma^2+2np-p^2.
 \label{eq:extPD}
\end{equation}
Thus the extremal daughter belongs locally to the
$\epsilon_2=-1$ PD sector, with an AdS$_2$-type longitudinal
geometry. 

\subsection{Rotating Nariai branch}
For the Nariai coincidence \eqref{eq:nar}, a fractional-linear
transverse coordinate gives the required canonical form. Set
\begin{equation}
 t_0=r_0-m>0,\quad s=a+\ell,\quad
 L=\frac{4t_0}{r_0},\quad K=s^2+t_0^2,
 \quad z=\frac{1+x}{1-x}.
 \label{eq:narconstants}
\end{equation}
We assume $C_f>0$, $m>0$, $r_0>m$, and $s\ne0$, and restrict the
coordinate identification to regular angular patches with $x\ne1$.
The condition $s\ne0$ is needed for the horizon-adapted Killing
coordinates used below, since $C_0=2s$. It is a restriction of this
coordinate construction, not a claim that the excluded parameters
admit no non-expanding limit. At the double root,
$D_0=2r_0t_0$ and $\lambda_h=-L$.
The exact A+ functions give
\begin{align}
 \frac{\rho_0^2}{(1-x)^2}&=U(z),
 &U(z)&=Kz^2+2mt_0z+m^2,\notag\\
 \frac{\Delta_x}{(1-x)^4}&=T(z),
 &T(z)&=z\left(\frac{t_0}{r_0}z+\frac{m}{r_0}\right).
 \label{eq:naridentities}
\end{align}
In particular, the discriminant of $U$ is $-4m^2s^2$. The function
$A(x)=1+\alpha^2(\ell^2-a^2)x^2$ in Eq.~\eqref{eq:ABCD}
is essential here; its cubic specialization $1-x^2$ cannot be used.
Equation~\eqref{eq:generaldouble} becomes
\begin{align}
 \dif s_{\mathrm N}^2={}&U(z)
 \left[LR^2\,\dif\tau^2-\frac{C_f}{L}\,\frac{\dif R^2}{R^2}\right]
 +\frac{C_fU(z)}{4T(z)}\,\dif z^2+\frac{4s^2T(z)}{U(z)}
 (\dif\Phi-2mR\,\dif\tau)^2.
 \label{eq:narmetric}
\end{align}
To express the quadratic transverse functions in canonical PD
form, introduce the normalization and shift constants
\begin{equation}
 A_*=\sqrt{\frac{C_fK}{L}},\qquad
 B_*=\frac{A_*mt_0}{K}.
 \label{eq:narAB}
\end{equation}
The complete canonical dictionary is
\begin{align}
 p&=A_*z+B_*,&q&=R,
 &t&=\frac{L}{\sqrt{C_f}}\tau,
 &y&=-s\sqrt{\frac{L}{K}}\,\Phi,\notag\\
 \gamma&=ms\sqrt{\frac{C_f}{LK}},
 &n&=\frac{2A_*m}{Lr_0}-B_*,
 &\epsilon_0&=0,
 &\epsilon_2&=+1.
 \label{eq:nardictionary}
\end{align}
Indeed, these quantities obey
\begin{align}
 p^2+\gamma^2&=\frac{C_f}{L}U(z),\notag\\
 P(p)=p^2+2np-\gamma^2&=\frac{4A_*^2}{L}T(z),\notag\\
 \dif y+2\gamma q\,\dif t
 &=-s\sqrt{\frac{L}{K}}(\dif\Phi-2mR\,\dif\tau).
 \label{eq:narmatching}
\end{align}
The Nariai daughter is therefore locally the canonical PD metric with
$Q(q)=-q^2$ and $\epsilon_2=+1$, whose longitudinal geometry is of dS$_2$ type. The real
coordinate dictionary in Eq.~\eqref{eq:nardictionary} is valid
under the stated nondegeneracy assumptions; in particular, the
case $a+\ell=0$ is not covered by this proof. On the Nariai locus
this exceptional case also satisfies $r_0=2m$ and is treated
separately below. This identification
concerns the formal double-root daughter; the global limit obtained
by approaching two separated horizons and the identification lattice
are separate questions.

\paragraph{Pure-NUT specialization.}
The Nariai correspondence also includes a nonrotating Kerr parameter
with nonzero NUT charge. 
For $a=0$, $m>0$, $\ell\ne0$, and $\alpha>0$ (hence
$0<\alpha|\ell|<1$ on the coincidence locus), the positive
black-hole--acceleration coincidence is a member of the generic
Nariai branch above. In this case
\begin{equation}
 r_0=m+\sqrt{m^2+\ell^2},\quad s=\ell,\quad
 K=m^2+2\ell^2,\quad
 L=\frac{4\sqrt{m^2+\ell^2}}{r_0}.
 \label{eq:tnnarparams}
\end{equation}
Consequently, the canonical parameter in
Eq.~\eqref{eq:nardictionary} specializes to
\begin{equation}
 \gamma_{\mathrm{TN}}=m\ell
 \sqrt{\frac{C_f}{L(m^2+2\ell^2)}}\ne0.
 \label{eq:tnnargamma}
\end{equation}
Thus the non-expanding pure-NUT double-root daughter belongs to
the $\epsilon_2=+1$ sector with a nonzero canonical fibration
parameter, not to the flat-longitudinal swirling sector.

\subsection{Exceptional Nariai endpoint}
The Nariai dictionary of Sec.~5.4 becomes singular at
$s=a+\ell=0$. On the Nariai locus this implies $r_0=2m$, $\alpha=(2m)^{-1}$ and
$a=-\ell$. At the coincident horizon the A+ functions instead obey
\begin{equation}
 C_0=C'_0=0,\qquad D_0=4m^2,\qquad
 \rho_0^2=4m^2,\qquad \Omega_0=1-x.
 \label{eq:narexceptiondata}
\end{equation}
The horizon generator is now $\partial_t$. We take
$t=D_0\tau/\varepsilon$, $\phi=\Phi$ and
$r=r_0+\varepsilon R$, with the same local double-root
scaling as above. Setting $z=(1+x)/(1-x)$, the limiting metric is
\begin{align}
 \dif s^2={}&U\left[2R^2\,\dif\tau^2
 -\frac{C_f}{2}\,\frac{\dif R^2}{R^2}\right]
 +\frac{C_fU}{4T}\,\dif z^2
 +\frac{16m^4T}{U}\,\dif\Phi^2,\notag\\
 &U=m^2(z+1)^2,\qquad T=\tfrac12z(z+1).
 \label{eq:narexceptionmetric}
\end{align}
With $A_0=m\sqrt{C_f/2}$, the canonical coordinates
and parameters are
\begin{equation}
\begin{alignedat}{4}
p&=A_0(z+1),\qquad&
q&=R,\qquad&
t&=\frac{2\tau}{\sqrt{C_f}},\qquad&
y&=2\sqrt{2}\,m\Phi,\\
\gamma&=0,\qquad&
n&=-\frac{A_0}{2},\qquad&
\epsilon_0&=0,\qquad&
\epsilon_2&=+1.
\end{alignedat}
\label{eq:narexceptionmap}
\end{equation}
Indeed, $p^2=(C_f/2)U$ and
$P(p)=p^2+2np=A_0^2z(z+1)=2A_0^2T$;
the fibre coefficient agrees as well. The exceptional endpoint therefore belongs to the same $\epsilon_2=+1$
sector, with $\gamma=0$. The chart is distinct from the $s\to0$
limit of the generic chart at fixed $\Phi$; periodic Killing
identifications remain unspecified.

\subsection{Cubic branch}
For the cubic coincidence $r_-=r_+=r_A^+$, the daughter is the
swirling metric of Eq.~\eqref{eq:swirldaughter}. Its canonical
PD representation is obtained by setting
\begin{equation}
 \epsilon_0=1,\qquad \epsilon_2=0,\qquad
 \gamma=1,\qquad n=2j,
 \label{eq:cubicpddataparam}
\end{equation}
with the coordinate transformation
\begin{equation}
 p=j\varrho^2,\qquad q=Z,\qquad
 t=T,\qquad y=\frac{\varphi}{2j}.
 \label{eq:cubicpddatacoord}
\end{equation}
The structure functions then become $Q(q)=1$ and $P(p)=4jp$.
Substitution into Eq.~\eqref{eq:npd} reproduces the complete
swirling metric of Eq.~\eqref{eq:swirldaughter}, establishing the
local identification with the $\epsilon_2=0$ non-expanding PD sector.
This coordinate dictionary applies away from the rotational axis;
its global angular identification is not fixed by the local matching.

\section{Near-critical geometry and the non-expanding contraction}\label{sec:nearcritical}
The homothety of Sec.~\ref{sec:universal} holds at exact cubic
criticality. To examine nearby configurations, write the three
approaching horizons and the radial coordinate as
\begin{equation}
 r_\pm=m(1\pm\varepsilon\delta),\qquad
 r_A^+=m(1+\varepsilon\Delta),\qquad
 r=m(1+\varepsilon R),
 \label{eq:nearcriticalsplit}
\end{equation}
where $m>0$, $\ell$, $\delta$, $\Delta$ and $R$ remain fixed as
$\varepsilon\to0$. We take $C_f$ to have a smooth positive limit
equal to its chosen normalization on the cubic-critical locus. The constants $\delta$ and $\Delta$ describe,
respectively, the splitting of the black-hole horizons and the
separation of the positive acceleration horizon from $r=m$. The corresponding
physical parameters obey
\begin{equation}
 a^2=m^2+\ell^2-m^2\varepsilon^2\delta^2,\qquad
 \alpha=\frac{1}{m(1+\varepsilon\Delta)}.
 \label{eq:nearcriticalparameters}
\end{equation}
The exact radial function and the north-end angular opening are
\begin{align}
 \Delta_r={}&-m^2\varepsilon^3(R^2-\delta^2)(R-\Delta)
 \frac{2+\varepsilon(R+\Delta)}{(1+\varepsilon\Delta)^2},
 \label{eq:nearcriticalradial}\\
 \left.\frac{\Delta_x}{1-x^2}\right|_{x=1}
 ={}&\frac{\varepsilon^2(\Delta^2-\delta^2)}
 {(1+\varepsilon\Delta)^2}.
 \label{eq:nearcriticalopening}
\end{align}
These identities hold for arbitrary finite $\ell$. The curves
$\Delta=\pm\delta$ describe the neighbouring double coincidences.
Since the stationary chart is singular at the radial horizons,
we introduce principal-ray coordinates
\begin{equation}
 \dif t=\dif v+\frac{\sqrt{C_f}D}{\Delta_r}\,\dif r,
 \qquad
 \dif\phi=\dif\psi-\frac{\sqrt{C_f}C}{\Delta_r}\,\dif r.
 \label{eq:nearcriticalingoing}
\end{equation}
Here $A,B,C,D$ denote the A+ functions introduced in
Sec.~\ref{sec:roots}. The identity $AD+BC=\rho^2$ cancels the
apparent radial pole, yielding the exact line element
\begin{align}
 \dif s^2=\frac{1}{\Omega^2}\bigg[{}&
 -\frac{\Delta_r}{\rho^2}(A\,\dif v-B\,\dif\psi)^2
 -2\sqrt{C_f}(A\,\dif v-B\,\dif\psi)\,\dif r \notag\\
 &\quad{}+\frac{C_f\rho^2}{\Delta_x}\,\dif x^2
 +\frac{\Delta_x}{\rho^2}(C\,\dif v+D\,\dif\psi)^2
 \bigg].\label{eq:nearcriticalregular}
\end{align}
A finite mixed radial term and rotational one-form are obtained with
the scaling
\begin{equation}
 v=\Phi,\qquad \psi=\frac{V}{\varepsilon},\qquad
 r=m(1+\varepsilon R).
 \label{eq:nearcriticalblowup}
\end{equation}
With $a_0=\sqrt{m^2+\ell^2}$, define
\begin{equation}
 \begin{gathered}
 B_0(x)=a_0(1+x^2)+2\ell x,\qquad
 C_0=2(a_0+\ell),\\
 \Sigma_0(x)=C_0B_0(x),\qquad
 P_0(x)=(1-x)^3(1+x).
 \end{gathered}
 \label{eq:nearcriticalleadingfunctions}
\end{equation}
The source functions obey
\begin{equation}
 \begin{gathered}
 A=1-x^2+O(\varepsilon),\quad B=B_0+O(\varepsilon^2),
 \quad C=C_0+O(\varepsilon),\\
 D=2m^2\varepsilon R+O(\varepsilon^2),\quad
 \rho^2=\Sigma_0+O(\varepsilon),\quad
 \Omega=1-x+O(\varepsilon).
 \end{gathered}
 \label{eq:nearcriticalleadingorders}
\end{equation}
Under this scaling, $A\,\dif v-B\,\dif\psi=O(\varepsilon^{-1})$
and $\Delta_r=O(\varepsilon^3)$. Their squared contribution to Eq.~\eqref{eq:nearcriticalregular}
is therefore $O(\varepsilon)$, whereas the mixed radial, angular
and rotational terms have finite limits. We obtain
\begin{align}
 \dif s_{\mathrm{lead}}^2=\frac{1}{(1-x)^2}\bigg[{}&
 2m\sqrt{C_f}\,B_0(x)\,\dif V\,\dif R
 +\frac{C_f\Sigma_0(x)}{P_0(x)}\,\dif x^2+\frac{P_0(x)}{\Sigma_0(x)}
 \big(C_0\,\dif\Phi+2m^2R\,\dif V\big)^2\bigg].
 \label{eq:nearcriticalfullmetric}
\end{align}
The parameters $\delta$ and $\Delta$ drop out of the leading
metric. The same expression follows by imposing cubic criticality
before the scaling. The homothety of Sec.~\ref{sec:universal}
therefore identifies this geometry locally with the swirling universe,
up to a constant scale, for any finite NUT rapidity and the stated
$C_f$ normalization.

The limit holds for fixed finite $\delta$ and $\Delta$ on regular
patches where $x\ne1$ and $\Sigma_0P_0\ne0$, assuming a smooth
positive limit of $C_f$. The approach parameters may reappear at
subleading order; the calculation does not establish uniformity at
the noncompact angular end.

The limit does not retain the negative acceleration horizon or the
parent curvature singularity. These features belong to the critical
parent geometry considered in Sec.~\ref{sec:geometry}, not to its
near-horizon daughter.

\section{Discussion and conclusions}\label{sec:conclusions}
The main result is the local homothety of the complete cubic-critical
accelerating Kerr--NUT metrics at finite NUT rapidity. With the
transverse normalization in Eq.~\eqref{eq:cubicCf}, the projective
transformation relates every member of this critical family to the
zero-NUT representative, up to a constant scale. Thus the NUT
rapidity does not furnish an independent local homothety modulus on
the cubic-critical locus. The relation holds for the critical parent
metrics, before any near-horizon limit is taken.

Barrientos and Cisterna identified the swirling universe as a
triple-horizon limit of the rotating C-metric~\cite{BC2026}. The
parent homothety shows that the corresponding cubic near-horizon
geometry remains locally homothetic to the swirling universe at
arbitrary finite NUT rapidity. It also relates the local geometry of
the critical horizons. In particular, a spatial section of the cubic
horizon has one finite axis and a noncompact angular end. The
single finite-axis elementary-flatness condition is compatible with
the homothety, but it does not determine the global Killing
identifications.

The other critical limits fit into the same non-expanding
Pleba\'nski--Demia\'nski framework. Explicit four-dimensional
coordinate dictionaries identify the ordinary extremal, Nariai and
cubic daughters with the negative-, positive- and zero-curvature
longitudinal sectors, respectively. The exceptional Nariai endpoint
requires a separate Killing chart but remains in the positive-curvature
sector. The near-critical calculation provides a complementary result:
for the specified cubic-adapted scaling, fixed first-order separations
of the approaching horizons do not affect the leading local metric.
This does not imply independence of the subleading corrections or of
other scaling procedures.

These conclusions concern local geometry. Neither the homothety nor
the elementary-flatness condition determines a globally regular
Killing quotient or a maximal analytic extension of the critical
parent. The behaviour of the near-critical limit at the noncompact
angular end and the possible commutation of the two double-root
limits at cubic criticality also remain open.

\end{document}